%% file: main.tex
\documentclass[sigconf]{acmart}

\AtBeginDocument{%
  }

\setcopyright{rightsretained}
\acmDOI{}
\acmConference[CHI '25 Workshop on Envisioning the Future of Interactive Health]{}{April 27th, 2025}{Yokohama, Japan}
\acmISBN{}
\copyrightyear{2025}
\acmYear{2025}

\usepackage{graphicx}
\usepackage{url}   

\newif\ifcomment
\commenttrue   
\ifcomment
  \newcommand{\missing}[1]{\textcolor{red}{~#1}}
  \newcommand{\kel}[1]{\textcolor{olive}{~Kellie: #1}}
  \newcommand{\ken}[1]{\textcolor{magenta}{~Kenny: #1}}
\else
  \newcommand{\missing}[1]{}
  \newcommand{\kel}[1]{}
  \newcommand{\ken}[1]{}
\fi

\begin{document}

\title{Envisioning an AI-Enhanced Mental Health Ecosystem}


\author{Kellie Yu Hui Sim}
\email{kelliesyhh@gmail.com}
\orcid{0009-0005-6451-7089}
\affiliation{
  \institution{Singapore University of Technology and Design}
  \country{Singapore}
}

\author{Kenny Tsu Wei Choo}
\email{kennytwchoo@gmail.com}
\orcid{0000-0003-3845-9143}
\affiliation{
  \institution{Singapore University of Technology and Design}
  \country{Singapore}
}

\renewcommand{\shortauthors}{Sim et al.}

\begin{abstract}
    \input{sections/0_abstract}
\end{abstract}



\maketitle

\input{sections/1_introduction}
\input{sections/2_applications}
\input{sections/3_overview}
\input{sections/4_discussion}
\input{sections/5_conclusion}

\bibliographystyle{ACM-Reference-Format} 
\bibliography{main}

\end{document}

%% file: sections/0_abstract.tex
The rapid advancement of Large Language Models (LLMs), reasoning models, and agentic AI approaches coincides with a growing global mental health crisis, where increasing demand has not translated into adequate access to professional support, particularly for underserved populations. 
This presents a unique opportunity for AI to complement human-led interventions, offering scalable and context-aware support while preserving human connection in this sensitive domain.
We explore various AI applications in peer support, self-help interventions, proactive monitoring, and data-driven insights, using a human-centred approach that ensures AI supports rather than replaces human interaction.
However, AI deployment in mental health fields presents challenges such as ethical concerns, transparency, privacy risks, and risks of over-reliance. 
We propose a hybrid ecosystem where where AI assists but does not replace human providers, emphasising responsible deployment and evaluation.
We also present some of our early work and findings in several of these AI applications.
Finally, we outline future research directions for refining AI-enhanced interventions while adhering to ethical and culturally sensitive guidelines.

%% file: sections/1_introduction.tex
\section{Introduction}
In recent years, mental health has emerged as a critical priority in our modern society, with an estimated 1 in 8 people worldwide living with a mental health condition~\cite{MentalDisorders2022a}, amplified by global shifts and increased awareness of well-being. 
Despite increasing awareness, many individuals still experience limited access to professional support~\cite{olearySuddenlyWeGot2018}, highlighting the need for innovative, scalable approaches that extend beyond traditional care settings. 

Recent AI advancements, including developments in LLMs, reasoning frameworks~\cite{wangCanChatGPTDefend2023}, and agentic approaches~\cite{qiuLLMbasedAgenticSystems2024}, offer a unique opportunity to reimagine mental health support. 
This is especially promising in the wake of the development of multimodal capabilities that integrate text, voice, images, and sensor data. 
For example, studies have shown that chatbots like Woebot and Tess can significantly reduce symptoms of depression and anxiety~\cite{fitzpatrickDeliveringCognitiveBehavior2017, fulmerUsingPsychologicalArtificial2018}. 
By integrating AI into mental health services, it is possible to design personalised, data-driven interventions that harness the capabilities of human care providers and foster meaningful interactions. 
This integration is especially promising in applications such as peer support, self-help interventions, and proactive monitoring, where AI can serve as a facilitator for timely and sensitive support.
However, despite their potential, concerns regarding ethical AI deployment, user trust, and data privacy remain. 

Our vision is to create AI systems that serve as supportive collaborators in mental health care, complementing rather than replacing human expertise and emotional understanding. 
By ensuring that technology remains a supportive partner in the therapeutic process, we believe that this human-centred approach can respect the deeply sensitive nuances of mental health care and foster meaningful long-term engagements. 
Human providers such as mental health professionals, peer supporters, and caregivers bring irreplaceable qualities like empathy, cultural understanding, and nuanced judgment that AI currently cannot fully replicate.

In this paper, we explore possible ways in which recent AI innovations can be incorporated into mental health applications. 
We examine potential use cases where AI-enhanced tools can enhance self-help mechanisms, facilitate peer support, and provide valuable insights for early intervention, all while maintaining a commitment to empathetic, human-focused care. 
Additionally, we provide an overview of our early work and ongoing research directions.
Through this, we aim to foster an environment where AI enriches mental health care, ultimately contributing to a more inclusive and responsive care ecosystem.

%% file: sections/2_applications.tex
\section{Exploring AI Applications in Mental Health}
AI is increasingly being integrated into mental health support, with applications spanning chatbots for therapy~\cite{fitzpatrickDeliveringCognitiveBehavior2017, fulmerUsingPsychologicalArtificial2018}, predictive analytics from social media data~\cite{radwanPredictiveAnalyticsMental2024}, providing guidance for supporters~\cite{hsuHelpingHelperSupporting2023}, and clinical decision support systems~\cite{tutunAIbasedDecisionSupport2023}. 
However, challenges remain in ensuring that these systems are transparent, user-friendly, and ethical~\cite{deneckeArtificialIntelligenceChatbots2021}.
A user-centred approach is essential for digital health interventions~\cite{thiemeDesigningHumancenteredAI2023a}, and we propose several paradigms that collectively present a multifaceted view of how AI can revolutionise mental health care. 

\paragraph{\textbf{AI-Simulated Client}}
AI can simulate realistic client behaviour, providing training scenarios for practising communication, assessment, and intervention strategies in a controlled environment, with opportunities for immediate feedback from trained professionals or AI-driven evaluation tools. 
Studies have demonstrated the use of AI-simulated clients in therapist training and AI-driven therapy research~\cite{wangClientCenteredAssessmentLLM2024, wangCanChatGPTDefend2023}, using structured interactions to mimic patient behaviour. 
In our early work featuring a web-based chat system, we implemented an LLM-simulated distressed client, which dynamically adapted to conversations and exhibited emotional states, engaging participants in structured peer support interactions and allowing them to experience varied emotional and behavioural responses.
A peer support setting was chosen as it can complement professional care by addressing emotional needs and providing practical guidance to those hesitant to seek formal help ~\cite{olearyDesignOpportunitiesMental2017, sharmaFacilitatingEmpathicConversations2021, shahModelingMotivationalInterviewing2022}. 
We assessed how effectively the client could simulate a believable peer support interaction, and found that the client could provide valuable training opportunities and exhibited certain human-like and realistic qualities.
However, further refinement is needed to balance human-likeness, realism, and adaptability, ensuring that simulated clients can capture the nuanced complexity of real-world interactions while providing sufficient variability to prepare users for diverse scenarios.

\paragraph{\textbf{AI as a Peer, Counsellor, or Therapist}}
AI-powered conversational agents (CAs) can provide empathetic, context-aware support for individuals in distress~\cite{liSystematicReviewMetaanalysis2023, schyffProvidingSelfLedMental2023, liuComPeerGenerativeConversational2024a}.
Leveraging LLMs and multimodal inputs, CAs can engage in dynamic and coherent conversations, responding to text, voice and even non-verbal cues, creating a more comprehensive and natural support experience similar to that of a human supporter. 
Studies have demonstrated that CAs such as Woebot and Tess can reduce symptoms of anxiety and depression among university students and adults~\cite{fitzpatrickDeliveringCognitiveBehavior2017, fulmerUsingPsychologicalArtificial2018}.
These AI-powered chatbots use cognitive behavioural therapy-based approaches to engage users in empathetic conversations while providing personalised coping strategies.
To be effective, users must clearly understand AI’s role, and interactions must balance guidance provided by AI with genuine human connection to prevent over-reliance while ensuring AI remains a reliable source of support and comfort.

\paragraph{\textbf{AI-Generated Suggestions}}
Beyond its role as a CA, AI can enhance real-time support sessions by analysing ongoing interactions and providing context-sensitive suggestions to both professional and peer supporters~\cite{hsuHelpingHelperSupporting2023, sharmaHumanAICollaboration2023}. 
In our study with the LLM-simulated client, participants were also presented with LLM-generated suggestions based on motivational interviewing techniques, empathetic responses, and informational or emotional support, assisting in challenging conversations while maintaining human oversight in the interactions.
This provided another perspective for supporters and offered them inputs, especially when they were unsure how to respond.
However, not all participants used the suggestions, with some highlighting that the suggestions should be advisory rather than directive, ensuring that they complement rather than disrupt the natural flow of interaction.
Others noted areas for improvement in the quality and adaptability of suggestions.
Clear indicators and rationales for AI-generated suggestions can help supporters maintain authenticity and preserve the human touch, while allowing human supporters to remain in control of the conversation.
Avoiding over-reliance on AI-generated suggestions is essential to preserving the human touch and connection defining effective support.

\paragraph{\textbf{AI for Decision-Support and Evaluation}}
AI reasoning models can assist clinicians in making informed decisions by offering diagnostic suggestions, risk assessments, and treatment options~\cite{tutunAIbasedDecisionSupport2023}. 
When used in a collaborative manner, these tools complement human judgement and contribute to improved clinical outcomes.
Explainable AI is crucial in this vision, and we envision that AI outputs can be presented in interpretable format that helps clinicians understand and trust the system’s recommendations. 
Integrating AI insights with human decision-making allows for bias monitoring, accountability and improved clinical outcomes.

\paragraph{\textbf{AI for Self-Help and Companionship}}
AI-powered self-help tools offer immediate, personalised mental health support, serving as an empathetic companion throughout one's mental health journey~\cite{liuUsingAIChatbots2022, schyffProvidingSelfLedMental2023}. 
These AI companions can engage in regular check-ins, offer emotional support, and personalised resource recommendations.
AI-driven self-tracking insights can support users by identifying patterns and suggesting coping strategies.
With intuitively-designed interfaces and sustained engagement mechanisms, we envision the fostering of long-term behavioural change through digital nudges.

\paragraph{\textbf{AI for Proactive Monitoring}}
By analysing data from digital sources such as wearable devices, mobile applications, social media, and more, AI can be used to detect early signs of mood changes or changes in emotional states, such as through changes in language use or behavioural patterns~\cite{adlerDetectionActionableSensing2024, xuMentalLLMLeveragingLarge2024a}. 
This proactive monitoring could potentially aid the detection of early warning signs through changes in digital behaviour patterns, enable timely interventions in such cases, and provide data-driven insights.
While promising, this approach must balance between its benefits, ethical considerations including user privacy, transparent user consent, and the design of feedback loops that effectively communicate findings. 

\paragraph{\textbf{AI as an Embedded and Ubiquitous Mental Health Companion}}
We are currently exploring how smart wearables, mobile applications, and AI can be seamlessly integrated into daily life for wellness tracking and mental health support.
By passively monitoring user interactions and physiological patterns, these systems can provide real-time, context-aware interventions or even nudges to encourage self-care practices~\cite{prajwalEfficientEmotionSelfreport2023, adlerDetectionActionableSensing2024}. 
We aim to collect data from participants through everyday digital interactions, complemented by smart wearables to track relevant physiological signals.
Participants will be asked to periodically report their emotional states, allowing for a dynamic and ecologically valid understanding of such states.
These solutions focus on low-resource, privacy-preserving architectures (e.g., edge computing) to ensure wide accessibility. 
Offline LLMs, such as Flan, can be integrated into these systems to provide immediate, contextually-relevant support while preserving user privacy.
The design of such environments should also serve users across various socio-economic and geographical contexts.
In everyday environments like homes, AI could play an ambient role, integrating seamlessly and subtly into users’ daily routines, supporting mental well-being passively and proactively.

%% file: sections/3_overview.tex
\section{Towards an AI-Enhanced Mental Health  Ecosystem}
While each proposed AI application addresses specific needs in the mental health space, there lies potential in their ability to work together as part of an interconnected ecosystem. 
A holistic ecosystem—one that is proactive, responsive, and adaptive—can enhance mental health care by seamlessly combining multiple AI paradigms.  
This section explores how these AI applications can complement one another to enhance mental health care.

\paragraph{\textbf{Leveraging Social Media and Online Forum Data}}
Grounding AI models in real-world mental health conversations can help to ensure context-aware, emotionally attuned responses. 
To achieve this, we are exploring the use of social media and online forum data as a baseline dataset, providing large-scale, real-world examples of support dynamics and help-seeking behaviours.
This dataset informs several of the applications highlighted earlier, allowing them to be more contextually aware, emotionally attuned, and aligned with real-world mental health conversations:
\begin{itemize}
    \item \textbf{AI-Simulated Client}: Training AI-simulated clients to replicate and capture the nuanced complexity of real user experiences.
    \item \textbf{AI as a Peer, Counsellor, or Therapist}: Enhancing CAs' ability to provide empathetic and contextually appropriate responses.
    \item \textbf{AI-Generated Suggestions}: Learning from effective supportive conversations to generate response recommendations that are empathetic and suitable.
    \item \textbf{AI as a Decision-Support and Evaluation Tool}: Improving AI-driven risk assessments using real-world help-seeking behaviours.
    \item \textbf{AI as a Proactive Digital Environment Monitor}: Detecting early signs of distress by analysing shifts in language patterns and behaviour.
\end{itemize}

\paragraph{\textbf{Combining Proactive Monitoring with AI Companions}}
A proactive AI monitor can detect subtle emotional shifts, triggering timely intervention. 
When distress signals arise, AI companions can check in with the user, suggest coping strategies or escalate support by recommending suitable human support services.
This integration ensures that AI not only reacts to crises but actively works to identify and address concerns before they become critical. 
However, balancing privacy, transparency, and user consent is essential to ensure ethical implementation.

\paragraph{\textbf{Enhancing Human-AI Collaboration}}
AI-generated suggestions and CAs can work together to enhance both human-supported and AI-supported interactions.
For human supporters, AI-generated suggestions can provide context-sensitive guidance, helping peer supporters and professionals maintain authenticity while structuring their responses. 
We envision a hybrid support setting where a CA interacts with users but human supporters oversee or intervene as needed. 
In such scenarios, the AI-generated suggestions can assist supporters in refining or contextualising responses before they are delivered. 
This integration ensures that AI remains adaptable and aligned with best practices, particularly in cases where nuanced judgment is required.
This can benefit novice supporters, helping them to navigate complex conversations while ensuring interactions remain human-centred and emotionally intelligent.

\paragraph{\textbf{AI-Simulated Clients and Decision Support in Supporter Training}}
AI-simulated clients combined with decision-support tools can provide realistic and interactive training for supporters in a controlled environment. 
While the former adapts to various responses, the latter can analyse interactions and offer feedback on communication, assessment, and intervention strategies, helping supporters refine their skills before working with real clients. 
There could even be an integration of AI-generated suggestions that offer alternative phrasing, question prompts, and response strategies.
This makes training more interactive, data-informed, and reflective of real-world complexities.

\paragraph{\textbf{Seamless AI Support in Everyday Life}}
Beyond crisis detection, embedded AI companions in wearables, mobile applications, and smart home assistants can provide continuous and non-intrusive support.
Such embedded AI companions could encourage self-care through a mobile application, while smart home assistants could suggest relaxation techniques and personalise interventions.
CAs could also complement these efforts by providing structured interventions when users require deeper support.
This integrated approach seamlessly embeds support into daily routines, fostering long-term well-being through habit formation, personalised nudges, and non-intrusive interventions.

%% file: sections/4_discussion.tex
\section{Discussion}
Our proposed paradigms highlight a multifaceted vision for AI-enhanced mental health support, spanning proactive monitoring, real-time interaction, training, clinical decision support, and continuous, everyday care. 
A critical element underlying many of these paradigms is the performance of LLMs, which serve as the engine driving most applications in our vision.

\subsection{Evaluating LLMs for Mental Health Applications}
Given their central role in our vision for AI-enhanced mental health support, it is essential to establish robust evaluation criteria for LLMs. 
We propose several key factors for comparison:
\begin{itemize}
    \item \textbf{Contextual Understanding}: Evaluates the ability to maintain coherent conversations over time~\cite{zhuCanLargeLanguage2024}.
    \item \textbf{Explainability and Transparency}: Measures how justified outputs are, fostering trust and informed decision-making~\cite{cambriaXAIMeetsLLMs2024}.
    \item \textbf{Ethical Performance}: Assesses response biases and the safeguards needed to prevent potential harm.
    \item \textbf{User Experience Impact}: Captures satisfaction, clarity, and emotional resonance to ensure interactions feel human-like and realistic.
    \item \textbf{Technical Performance}: Examines latency, scalability, and real-time adaptability, critical for time-sensitive applications.
    \item \textbf{Cultural Understanding}: Tests sensitivity to cultural nuances, ensuring context-aware and appropriate responses. 
\end{itemize}

These criteria should be evaluated through user studies, expert assessments, and real-world implementations to ensure that AI models align with clinical and ethical expectations.

\subsection{Challenges and Ethical Considerations}
As AI applications become more integrated into mental health care, ethical, transparency, and accessibility concerns must remain central to their design.  
Key challenges include:
\begin{itemize}
    \item \textbf{Transparency and User Understanding}: Users must have clear visibility into how AI produces its insights. Mental health interventions require trust, and non-transparent reasoning from AI could undermine user confidence.
    \item \textbf{Privacy and Data Protection}: For applications requiring continuous data collection, guidelines should govern how data is stored, processed, and shared.  
    \item \textbf{Human Oversight and Decision-Making}: While AI can enhance efficiency, it should not replace human judgement, especially in sensitive mental health contexts. AI-generated insights must remain advisory rather than directive, ensuring that users can retain the final decision-making authority.  
    \item \textbf{Cultural and Socioeconomic Accessibility}: AI-enhanced mental health systems must be inclusive, ensuring language adaptability, culturally sensitive responses, and accessibility in low-resource settings. Offline models and edge computing solutions could help to reach underserved populations.  
\end{itemize}

\subsection{Future Directions}
While LLMs provide fluent and engaging conversations, they often lack precision in task-specific interventions.
In contrast, structured reasoning models that outline their thought processes can offer greater interpretability and accuracy in decision-making.
A hybrid approach that integrates LLMs with structured reasoning models could enhance both reliability and trust in AI-enhanced mental health applications.
To ensure these systems generate context-sensitive and interpretable outputs, expert validation before deployment remains essential.

For AI-enhanced mental health applications to be effective at scale, they must be continuously refined through iterative user studies and interdisciplinary collaboration. 
Pilot implementations in clinical and peer support settings should be done to assess real-world effectiveness, while long-term impact on users' well-being should be assessed through longitudinal studies.

Establishing ethical AI governance frameworks will be crucial to ensuring accountability and fairness.  
Hybrid AI models that integrate LLMs with structured reasoning tools should be explored further to enhance accuracy and transparency.  
Cross-cultural validation is also necessary to ensure AI applications remain sensitive to diverse populations and mental health paradigms.

Given that LLMs are still prone to hallucinations and unpredictable outputs~\cite{huangSurveyHallucinationLarge2025}, ongoing monitoring and improvements are essential.  
Striking the right balance between automation and human oversight will remain a critical challenge in designing AI that is both effective and trustworthy.  

By addressing these challenges and integrating insights from research in fields such as human-computer interaction and clinical psychology, AI-enhanced mental health systems can move towards safer, more equitable, and more effective deployment.

%% file: sections/5_conclusion.tex
\section{Conclusion}
In this work, we explore the transformative potential of AI in the mental health space, spanning diverse applications from peer support to ubiquitous AI companions. 
By integrating AI into mental health care, we highlight opportunities for enhanced accessibility, proactive intervention, and collaborative AI-human engagement. 
Additionally, evaluating different AI models presents a promising avenue for identifying the most effective frameworks for specific mental health applications.
As AI-enhanced interventions continue to evolve, it is crucial to prioritise rigorous evaluation, cultural sensitivity, and ethical considerations. 
Striking a balance between technological innovation and human connection remains fundamental to ensuring that AI serves as a responsible and effective tool in mental health care. 
Future research should focus on refining AI applications through interdisciplinary collaboration, real-world testing, and adaptive frameworks that meet the diverse needs of global populations.